\documentclass[12pt]{article}
\usepackage[margin=3cm]{geometry}
\usepackage{amsmath}
\usepackage{graphicx}

\begin{document}

%%%%%%%%%%%%%

\thispagestyle{empty}

\vspace*{0.3in}

\begin{center}
{\large \bf Towards a Holographic Description of Inflation
and
\\[.1in]
 Generation of Fluctuations from Thermodynamics}

\vspace*{0.7in} {Yi Wang}
\\[.2in]
{\em
Physics Department, McGill University,\\ Montreal, H3A2T8, Canada
\vspace{0.7in} }
\end{center}

\begin{center}
{\bf Abstract}
\end{center}
\noindent

Recently, Verlinde conjectured that gravity may be an entropic force, arising from thermodynamics on the holographic screen. We investigate the implications of the entropic force conjecture for inflationary cosmology. We find the background dynamics of inflation can be dually described in the holographic language. At the perturbation level, two kinds of novel scale invariant scalar fluctuations arise from thermal fluctuations on the holographic screen. These fluctuations can be responsible for CMB anisotropy and structure formation.

\vfill

\newpage

\setcounter{page}{1}

%%%%%%%%%%%%%

\section{Introduction}

The inflationary cosmology \cite{Guth:1980zm} has become the most promising candidate for the dynamics of the very early universe. At the background level, inflation can solve the flatness, horizon and monopole problems of the standard hot big bang cosmology. At the perturbation level, the quantum fluctuation generated during inflation provides seeds for the CMB anisotropy \cite{CMBobserve} and the large scale structure \cite{Mukhanov81}.

In this paper, inspired by Verlinde's entropic force conjecture \cite{Verlinde:2010hp}, we propose a holographic description of inflation. Before discussing inflation, we would like to first briefly review Verlinde's entropic force conjecture.

As early as the discovery of black hole thermodynamics, people start to realize the similarity between Einstein's equation and thermodynamics \cite{Bardeen:1973gs}. Based on these progresses, Jacobson \cite{Jacobson:1995ab} proposed that Einstein's equation can be derived as an equation of state near the Unruh horizon. Subsequently, this issue is also discussed by many authors. Especially, Padmanabhan \cite{Pad1} discussed the equipartition rule and thermodynamical aspects of gravity. Recently, Verlinde \cite{Verlinde:2010hp} discussed this issue in a more intuitive way, and conjectured that the origin of gravity is an entropic force arising from the thermodynamics on holographic screens.

Verlinde postulated that when a test particle with mass $m$ is located near a holographic screen, with distance $\Delta x$, the change of entropy on the holographic screen takes the form \footnote{For simplicity, we shall use the natural unit $\hbar=c=k_B=1$. We shall keep the Planck mass $M_p\equiv 1/\sqrt{8\pi G}$ explicit.}
\begin{equation}\label{eq:assumption}
  \Delta S=2\pi m \Delta x~.
\end{equation}

When the entropy of a system depends on position $\Delta x$, an entropic force $F$ arises as thermodynamical conjugate of $\Delta x$, satisfying the first law of thermodynamics on the holographic screen
\begin{equation}\label{eq:firstlaw}
  F \Delta x=T \Delta S~.
\end{equation}

Inserting Eq. \eqref{eq:assumption} into Eq. \eqref{eq:firstlaw}, we have
\begin{equation}
  F=2\pi m T~.
\end{equation}

Thus once one obtains a temperature of the system, the entropic force can be calculated. If one considers an accelerating reference frame with Unruh temperature
\begin{equation}
  T=\frac{a}{2\pi}~, \qquad a\equiv d^2 x/dt^2~,
\end{equation}
Newton's second law $F=ma$ arises as a result of the holographic screen thermodynamics.

Alternatively, one can consider a source of energy $E$, which is represented on a spherical holographic screen with radius $R$. The temperature can be obtained from the assumption that the equipartation rule applies on the holographic screen
\begin{equation}
  E=\frac{1}{2}NT~, \qquad E=M~, \qquad N=A/G~, \qquad A=4\pi R^2~.
\end{equation}
From equipartation, Newtonian gravity $F=GMm/R^2$ is recovered.

Since the entropic force conjecture is proposed, there have been several works investing various aspects of physics related to this conjecture. For example, the dynamics of apparent horizon \cite{Shu}, the Friedman equation \cite{Pad}, results on loop quantum gravity \cite{Smolin}, holographic actions for black hole entropy \cite{Caravelli:2010be}, UV/IR relations, holographic dark energy and an improved intuitive holographic screen \cite {Li:2010cj} are obtained from the entropic force conjecture.

The most inspiring view point in Verlinde's formalism, which is critically used in the present paper is that gravity is conjectured to be an entropic force. Thus the current understanding of gravity is only valid in the thermodynamical limit. However, in the early universe, when the universe has extremely high energy density, a holographic screen, which is of the size of the Hubble scale, might not have as many degrees of freedom as to have a good thermodynamical limit. In this case, the thermal fluctuations on the holographic screen induce fluctuations of gravity. The induced gravitational fluctuations become novel sources of density fluctuations. That will lead to the main result of the present paper.

In the present paper, our analysis is based on two assumptions: Firstly, the inflationary universe can be described by the dynamics of a holographic screen. This has to be true because the statement of holographic principle. Secondly, gravity is an entropic force arises from the thermodynamics on the holographic screen. This is the key observation of Verlinde's formalism. We have not used the details of Verlinde's entropic force formalism. \footnote{Verlinde's formalism have to be improved in order to describe the acceleration of the universe. This is because to derive General Relativity, one has to use the Komar mass, which is negative for the matter component that drives the acceleration of the universe. To avoid this problem, two approaches have been available at the moment: one is to add the future event horizon as a second holographic screen \cite{Li:2010cj}. Another is to give up some details of Verlinde's approach and start from more general principles, which is present here. We thank Miao Li to point this out \cite{miao}. }

In Section \ref{sec:bg}, we set up the background dynamics of holographic inflation. The detailed dynamics on the holographic screen is not known so far. However, inspired by Verlinde's equipartation rule, we argue that one can construct a number of toy models on the holographic screen which describes inflation with a graceful exit. We also propose an explicit toy model which can describe inflation, end of inflation, and dark energy holographically.

In Section \ref{sec:pert}, we discuss cosmic perturbation theory in the holographic inflationary framework. We find two or three sources of density fluctuations, either might dominate the curvature perturbation. One possible source is the perturbation of the inflaton field, if inflaton exists, which is a well known candidate for cosmic perturbations. However, two other kinds of perturbations also arise, one is the thermodynamical perturbation of the thermalized system, and another is the thermodynamical perturbation on the rate of thermalization.

We shall show that the amplitude of the thermal perturbations of the thermalized degrees of freedom on the holographic screen is of the same order as gravitational waves. Thus whenever an inflaton field exists, this thermal perturbation will be always subdominant. However, the other kind of perturbation, corresponding to the fluctuation of the thermalization rate, will be able to dominate the cosmic perturbations in some parameter regions.

Alternatively, one might be able to consider the case that inflation is purely described by holography without an inflaton. In this case, either of the two kinds of thermal perturbations could dominant the density perturbation, depending on parameter choices.

\section{The background dynamics}\label{sec:bg}

In this section, we shall setup the homogeneous isotropic background dynamics for inflation on the holographic screen. The holographic screen is naturally set to be a sphere surrounding the observable universe during inflation. The dynamics could be thought of in two ways: one can either identify the background described on the holographic screen as a dual description of an inflaton field, or think of the holographic screen as a unique description of the system, which do not have semi-classical counterpart as inflaton. As holography is not so far transparent to us, we shall leave both of the above possibilities open.

As we only have limited knowledge on holographic screens, before building explicit models, we would like to first discuss some general principles needed for a holographic description of inflation.

Firstly, one need to have inflation.\footnote{There exist several alternatives of inflation, which are also consistent with current observations. The possibility of holographic universe without inflation will be discussed in a separate publication \cite{prep}.} Inflation can be described by a Hubble size holographic screen with nearly constant degrees of freedom (thus nearly constant area).

Secondly, one need to have a graceful exit for inflation. There are several possibilities to terminate inflation. One possibility is a local tunneling from a false vacuum to a true vacuum. However, as in usual inflation models, this possibility can not directly lead to the current observable stage of inflation. This is because during inflation, different local patches of the universe do not have casual contact. Thus different patches of the universe reheats at hugely different time, resulting in eternal inflation and also too large density perturbations.

In the semiclassical description of inflation, another possibility which can terminate inflation is that inflation is driven by an inflaton field, which rolls down its potential towards an end of inflation. Inspired by this picture, one can try to find either holographic descriptions for inflaton, or holographic alternative for inflaton, which leads to inflation and graceful exit. In this section, we shall discuss holographic descriptions of inflation along this way. We are not aware of whether we are working on a holographic description of inflation, or a holographic alternative of inflaton. Because in Verlinde's framework, we do not have a precise dictionary for the bulk/boundary duality.

On the holographic screen, there are various possibilities to describe the holographic dynamics that leads to a termination of inflation. For example, one could have an inflaton field, and let it roll explicitly; one could add nonlinear corrections to Eq. \eqref{eq:assumption} to have rolling background; one could modify the equipartation rule to have modified gravity; one could assume non-equilibrium thermodynamics on the holographic screen to have entropy increase; one could effectively add some entropy source on the holographic screen; one could use various kinds of horizons for inflation instead of the Hubble horizon to be the holographic screen; and so on and so forth.

In the reminder of this section, we shall discuss a simple toy model of inflation described on the holographic screen. Note that this description is only for illustrating purpose. Because as we have noticed, there are a number of other toy models. All these models can be analyzed using similar methods as in this paper, especially for the mechanism to generate density perturbations.

As a toy model, we postulate that there are two species of degree of freedom, namely $A$ and $B$ on the Hubble size holographic screen with weak interaction with each other. We further assume that at the beginning, $A$ is hot, while $B$ is cold. The number of degrees of freedom $N_B$ is much greater than $N_A$.

One lesson we have learnt from Verlinde's entropic force is that space is no longer fundamental, but become emergent from the thermodynamics of the holographic screen. Thus when the $B$ particles are cold and not yet thermalized, they do not contribute to the emergence of space. The space volume inside the holographic screen only knows about $B$ when the $B$ particles are thermalized.

One should note that here ``thermalized'' is used in a loose sense. Because thermalization happens gradually, not suddenly. Anyway, as a toy model, we could assume that when particle $B$ reaches a temperature threshold $(T_A-T_B)/T_A<\Delta$, the $B$ particle is considered to be thermalized and can be felt by the emerged space. Here $\Delta$ is a parameter taking values from 0 to 1. The definition here might be over simplified. But we shall use this as a toy model anyway to illustrate the mechanism for generating fluctuations.

There are two further possibilities for the thermalization process: If the $B$ particle has very strong self interaction, there is a partial thermalization among $B$ particles. Otherwise, one part of the $B$ particles are thermalized with $A$ while the other part remains cold. (Of course, there are also intermediate cases lies in between these two possibilities.) We shall consider the former possibility in the remainder of the paper, because the former possibility leads to a more sharp end of inflation \footnote{However, we could add one more assumption that compared with the temperature of $A$, the particle $A$ is heavy and the particle $B$ is light. In this case, a intermediate case of the former and the latter can also lead to sharp end of inflation. This is because, in a semiclassical picture, at the beginning when the $B$ particle is cold, both $A$ and $B$ move slowly. After a period of thermalization, the speed of the $B$ particle approaches to the speed of light and the thermalization process is greatly enhanced.}, which seems more consistent with the current observations.

With the assumptions for thermalization rate and its relation with emerged space, now we are able to setup our background model of holographic inflation. Initially, the size of the holographic screen does not change, because $B$ is not thermalized and the space inside the holographic screen does not know the existence of $B$. A fixed size of holographic screen is dual to a quasi-de Sitter universe. This is because the area of cosmic horizon is kept fixed, which can be identified to be the de Sitter horizon.\footnote{Here by quasi-de Sitter, we do not mean the horizon area changes, but rather, we have a breaking of de Sitter time translation invariance due to the process of thermalization. This is similar to ghost inflation \cite{ArkaniHamed:2003uz}, where the Hubble parameter does not change, while inflation can end and the causal structure of the universe is not like de Sitter.} The Hubble parameter is emergent and can be identified to be of the same order as the size of the holographic screen: $H\sim \sqrt{M_p^2/N_A}$.

After a period of thermalization, where the length of time depends on the interaction rate between $A$ and $B$, $B$ become thermalized, and the emergent space become aware of the existence of $B$. The Hubble horizon suddenly increases, resulting an end of inflation. If we like, we can also assume that $B$ is responsible for dark energy, by letting $N_B\sim M_p^2/H_0^2$, where $H_0$ is the Hubble parameter in the present time.

After the end of inflation, reheating follows. In this toy model, reheating might take place due to gravitational creation of particles. Or alternatively, reheating might also come from some bulk dynamics which is not described explicitly in the toy model, such as a decay and release of vacuum energy for the inflaton field. We shall not discuss reheating of the universe in detail here. Rather assuming the existence of reheating. In this case, after reheating, the universe enters radiation dominated phase and matter dominated phases. After that, when Hubble radius of the universe becomes close to $\sqrt{M_p^2/N_B}$, the universe becomes dark energy dominated.

\section{Nearly scale invariant perturbations}\label{sec:pert}

In this section, we discuss the cosmic perturbation theory in the entropic force framework. One might think that as in our toy model the Hubble parameter does not change during inflation, the density perturbation is ill defined. However, it is not the case because the time translation invariance is broken by the thermalization process. The time delay formula $\delta\rho/\rho \sim H\delta t$ can be used to calculate the density perturbation, where $\delta \rho$ is defined on a flat slice when the perturbation reenter the horizon.

There are two or three kinds of scalar perturbations in our toy model. One possible type of perturbation arises from the inflaton field, if there is an inflaton in the model. If there is an inflation field $\phi$, which defines a reheating surface at constant $\phi$, then the scalar field behaves as a source of density perturbation, with
\begin{equation}
  \left(\frac{\delta\rho}{\rho}\right)_\phi = H(\delta t)_\phi = H \frac{\Delta \phi}{\dot\phi} = \frac{H^2}{2\pi \dot\phi}~,
\end{equation}
where $\Delta \phi$ denotes the quantum fluctuation of $\phi$ at horizon exit \footnote{Here we still use the terminology ``horizon exit''. In a holographic description, this might be not accurate as the information ``outside'' the horizon is actually represent by the horizon. However, from the complementary principle, one can alternatively use the ``horizon'' view point safely.} averaged in the Hubble volume, satisfying $\Delta\phi=H/(2\pi)$. This amplitude is the same as that in slow roll inflation models. As we have discussed in Section \ref{sec:bg}, depending on the detailed dictionary of holography, $\phi$, together with the perturbation of $\phi$, might either exist or not exist in the entropic force description for inflation.

Now we come to the novel part of perturbations, originated from the thermal fluctuations on the holographic screen, which is responsible for the fact that inflation happens at a high energy scale, and the number of degrees of freedom on the holographic screen is not huge.

One thing to note is that the thermal fluctuation of the $A$ particle induces a change of Newtonian potential at horizon exit. This is because from Verlinde's entropy force framework, the Newtonian potential is related to the ratio between the entropy in the bulk and the entropy on the holographic screen.
When there is thermal fluctuations for entropy on the holographic screen, there is fluctuations for the Newtonian potential, of order
\begin{equation}
  \Phi \sim \Delta S \sim \frac{1}{\sqrt{N_A}}~.
\end{equation}
The relation $\Delta S\sim 1/\sqrt{N_A}$ reflects the fact that the derivation from thermal equilibrium is suppressed by the square root of the particle number.

This Newtonian potential is conserved after horizon exit, which induces a density perturbation
\begin{equation}
  \left(\frac{\delta\rho}{\rho}\right)_A \sim \frac{1}{\sqrt{N_A}}~.
\end{equation}
Note that $N_A$ can be related to $H$ as $N_A \sim M_p^2/H^2$, we have
\begin{equation}
  \left(\frac{\delta\rho}{\rho}\right)_A \sim\frac{H}{M_p}~.
\end{equation}
Here the amplitude is similar to the amplitude of gravitational waves (as will be discussed later in this section). If there exists an inflaton field, the density fluctuation from the quantum fluctuation of the inflaton field is larger than $\left(\delta\rho/\rho\right)_A$. On the other hand, if there is no inflaton, $\left(\delta\rho/\rho\right)_A$ might dominate the density perturbations. In this case, the tensor to scalar ratio will be $r\sim {\cal O}(1)$. This is not favored by the current observations. However, as there is an order one constant undetermined, there is still chance for $\left(\delta\rho/\rho\right)_A$ to dominate and be consistent with current observations.

In this simplest toy model, the power spectrum induced by $A$ does not have tilt. Tilts can be introduced when there are some time evolution of $H$, which can be achieved by replacing the sharp thermalization by a gradual thermalization.

Another fact we should note is that the thermalization process is related to reheating, thus a fluctuation on the thermalization rate should be responsible for density fluctuations as well. The fluctuation on thermalization rate takes the form
\begin{equation}
  \left(\frac{\delta\rho}{\rho}\right)_B = H(\delta t)_B=H\frac{\Delta T_B}{dT_B/dt}~,
\end{equation}
where $\Delta T_B$ is the fluctuation of the temperature of $B$ particles from a deviation from thermal equilibrium. As statistical fluctuations,  $\Delta T_B$ can be estimated as
\begin{equation}
  \frac{\Delta T_B}{T_B}  \sim \frac{1}{\sqrt{ N_B}}~.
\end{equation}
Thus the density fluctuation of this kind takes the form
\begin{equation}
  \left(\frac{\delta\rho}{\rho}\right)_B \sim \frac{H}{\sqrt{N_B}~d(\log T_B)/dt}~.
\end{equation}
As expected, the amplitude of density fluctuation depends on the thermalization rate of the $B$ particle. Note that if we would like to let  $N_B$ to represent dark energy, $N_B$ should be extremely large: $N_B \sim 10^{120}$. In this case, in order to have contribution to density fluctuation of order $10^{-5}$, the thermalization rate should be extremely slow, namely $d(\log T_B)/dt \sim 10^{-58}H$. On the other hand, if one gives up the requirement for $N_B$ to describe dark energy, and let the cosmic horizon become large by some other mechanisms (for example, one more step of thermalization with $C$ particles), then $d(\log T_B)/dt$ do not need to be so small.

The spectral index takes the form
\begin{equation}
  (n_s-1)_B=-\frac{d^2 (\log T_B)/dt^2}{H d(\log T_B)/dt}~.
\end{equation}
The detailed dynamics of thermalization determines the tilt of the spectrum.

We also would like to comment a little bit about tensor perturbations in holographic inflation. As we are considering Newtonian approximation of gravity, our calculation does not account for gravitational waves. In Verlinde's entropy force formalism, Einstein gravity can also be derived. In the framework of Einstein gravity, gravitational waves can be explicitly investigated. We hope we have the chance to address this issue in detail in the future.

However, before an explicit analysis, we should expect that gravitational waves in holographic inflation does not directly have quantum origin, because in the entropic force formalism, gravity is emergent and do not need to be quantized separately. However, gravitational waves might be able to be quantized as phonons, which effectively reflect the quantum nature of the degrees of freedom on the holographic screen \cite{mli}. One can also expect that the thermal fluctuations on the holographic screen induce the fluctuations of gravitational waves. The power spectrum of gravitational waves from both these origins are expected to be
\begin{equation}
  P_T \sim \frac{H^2}{M_p^2}~.
\end{equation}
The tilde of gravitational wave power spectrum should be $n_T=-2\epsilon$. as usual, which account for the change of the Hubble radius. We note that in our toy model, $H$ do not change until inflation suddenly terminates. Thus $n_T=0$ during inflation.

Finally, we would like to make several comments on our discussion of cosmic perturbation theory in the entropic force framework.

Firstly, we emphasize that although the analysis in this section is made for the toy model proposed at the end of Section \ref{sec:bg}, the mechanism can be applied to more general models. One can expect that in other setups of entropic force formalism, the cosmic perturbations can be analyzed using the same method. The mechanism to generate perturbations, instead of the background toy model with several unjustified assumptions, is the main result of the paper.

Secondly, the calculation of density perturbation here is not entirely holographic. This is because we have to use the time delay formula, which is not derived holographically. However, we note that we should have some connection between the holographic description and our real world (even yet without a detailed dictionary), because holography hides deep inside the description of our real world, not lies apparently. The time delay formula serves as a good connection in this sense. We also used the semiclassical picture of Newtonian potential, which can also be related to the time delay in the framework of standard cosmic perturbation theory.

Thirdly, in the bulk point of view, the analysis we have made is based on a fluctuation mode at the time of horizon exit. To translate this formulation to a fluctuation depending on spatial coordinate $\mathbf x$ or wave number $\mathbf k$ is straightforward, which will not be reviewed here.

Fourthly, even if the inflaton field exists in the holographic description, the fluctuation $(\delta\rho/\rho)_A$ on the holographic screen discussed above is different from the fluctuation of the inflation field. This is because $(\delta\rho/\rho)_A$ arises as a departure of Newtonian (or Einstein) gravity, which will not show up if gravity is not entropic. The second kind of fluctuation $(\delta\rho/\rho)_B$ might be the same with the inflaton fluctuation. But again, it is too early to conclude before one has a dictionary between the holographic screen and the bulk.

Fifthly, we note that the thermal fluctuations on the holographic screen are nearly Gaussian. Thus the induced density fluctuation should also obey nearly Gaussian statistics. The non-Gaussian estimator $f_{NL}$ is typically of order one for thermal fluctuations \cite{Chen:2007gd}, which is larger than slow roll inflation, but still smaller than the observational sensitivity of the experiments in the near future. However, in \cite{Chen:2007gd}, we also noticed that there exist some cases that $f_{NL}$ can be enhanced. It is interesting to see whether similar mechanisms takes place on the holographic screen, and how the bulk fluctuations are affected.

\section{Conclusion}

To conclude, we proposed a holographic description of inflation inspired by Verlinde's entropic force formalism. We discussed various possibilities towards a realistic holographic description of inflation models, especially how can inflation have a graceful exit. We investigated one of these models in more detail, namely the model with thermalization between two kinds of particles.

We find that two novel kinds of density fluctuations arises from thermal fluctuations on the holographic screen. One of them originates from the deviation of Newtonian (or Einstein) gravity due to thermal fluctuations of the thermalized system. The other originates from thermal fluctuations for the thermalization rate.

The analysis in this paper is semi-quantitative, because we do not have a precise definition of holographic duality in cosmology. It should be interesting to postulate some more concrete correspondences between the bulk and the holographic screen, inspired by AdS/CFT or similar well defined holographic dualities. In this case, it is hopeful to use the results in this paper to make observational predictions and test the entropic force conjecture concretely. As the entropic force conjecture is equivalent with the Einstein equation at large scales, the deviation from thermodynamical laws in small scales should be a valuable test for the validity of the entropic force conjecture.

\section*{Acknowledgments}
 We would like to thank Robert Brandenberger, Miao Li and Chunshan Lin for discussion on the present work.
 We also thank Alejandra Castro, Andrew Frey, Alex Maloney, Omid Saremi, Marcus Tassler and Bret Underwood for discussion on the issue of entropic force. This research was supported by NSERC and an IPP postdoctoral fellowship.

\end{document}